\documentclass[preprint,showpacs,preprintnumbers,amsmath,amssymb]{revtex4-1}
\usepackage{graphics,epsfig,subfigure}
\usepackage{diagbox}
\usepackage[usenames]{color}
\usepackage[colorlinks,
            linkcolor=blue,
            anchorcolor=red,
            citecolor=red
            ]{hyperref}
\usepackage{color}
\usepackage{graphicx}
\usepackage{amsmath}
\usepackage{amsfonts}
\usepackage{amssymb}
\usepackage{txfonts}
\usepackage{indentfirst}
\usepackage{booktabs}

\begin{document}
\renewcommand{\baselinestretch}{1.3}
\newcommand\beq{\begin{equation}}
\newcommand\eeq{\end{equation}}
\newcommand\beqn{\begin{eqnarray}}
\newcommand\eeqn{\end{eqnarray}}
\newcommand\nn{\nonumber}
\newcommand\fc{\frac}
\newcommand\lt{\left}
\newcommand\rt{\right}
\newcommand\pt{\partial}

\title{Rotating hybrid axion-miniboson stars}
\author{  Yan-Bo Zeng, Hong-Bo Li, Shi-Xian Sun, Si-Yuan Cui and Yong-Qiang Wang\footnote{yqwang@lzu.edu.cn, corresponding author}
}

\affiliation{Lanzhou Center for Theoretical Physics, Key Laboratory of Theoretical Physics of Gansu Province, School of Physical Science and Technology, Lanzhou University, Lanzhou 730000, People's Republic of China\\
Institute of Theoretical Physics $\&$ Research Center of Gravitation,
Lanzhou University, Lanzhou 730000, People's Republic of China}

\begin{abstract}
  We construct rotating hybrid axion-miniboson stars (RHABSs), which are asymptotically flat, stationary, 
  axially symmetric solutions of (3+1)-dimensional Einstein-Klein-Gordon theory. 
  RHABSs consist of a axion field (ground state) and a free complex scalar field (first excited state). 
  The solutions of the RHABSs have two types of nodes, including $^1S^2S$ state and $^1S^2P$ state. 
  For different axion decay constants $f_a$,
  we present the mass $M$ of RHABSs as a function of the synchronized frequency $\omega$, as well as the nonsynchronized frequency $\omega_2$, 
  and explore the mass $M$ versus the angular momentum $J$ for the synchronized frequency $\omega$ and the nonsynchronized frequency $\omega_2$ respectively. 
  Furthermore, we study the effect of axion decay constant $f_a$ and scalar mass $\mu_2$ on the existence domain of the synchronized frequency $\omega$. 
\end{abstract}

\maketitle

\section{Introduction}\label{Sec1}

One of the most attractive problems in the cosmology is the nature of dark matter. 
Boson stars (BSs) are considered as a possible explanation for parts of the dark in the halo of galaxies~\cite{Sahni:1999qe,Hu:2000ke,Matos:2000ng}. 
In the 1950s, John Wheeler explored the classical electromagnetic field coupled to Einstein gravity and introduced the concept of geons~\cite{Wheeler:1955zz,Power:1957zz}. 
However, no stable geons are found in Einstein-maxwell theory. 
Later, Kaup~{\em et al.} constructed a complex scalar field coupled to Einstein gravity theory, which was called as Einstein--Klein--Gordon~(EKG) theory. 
BSs described as macroscopic Bose-Einstein condensates is a kind of solutions for Einstein-Klein-Gordon theory. 
Heisenberg uncertainty principle provides repulsive force against gravitational collapse, mainly. 
Moreover, there are not only the ground state solutions with nodeless scalar field, 
but also spherical excited state solutions in Refs.~\cite{Lee:1986ts,Friedberg:1986tp}. 
They focus on spherical symmetry configuration, and it is likewise meaningful to understand the rotating boson stars. 
The cases of rotation were studied by Schunck and Mielke~\cite{schunck1996rotating,Schunck:1996he}, Yoshida and Eriguchi~\cite{Yoshida:1997qf},
Herdeiro and Radu~\cite{Herdeiro:2015gia}. 
Furthermore, the rotating axisymmetric solutions of BSs were generalized to the excited state case~\cite{Collodel:2017biu,Wang:2018xhw,Herdeiro:2018djx}. 
See Refs.~\cite{Schunck:2003kk,Liebling:2012fv} for a review. 

Besides, Einstein gravity coupled to a single scalar field, it can also be coupled to several different fields. 
In Ref.~\cite{Deng:1998dx}, Deng and Huang dealt with the case of two scalar fields coupled to gravity, 
in which two different kinds of scalar particles coexist the ground state. 
Then, Bernal {\em et al.}~\cite{Bernal:2009zy} constructed the model which has two states, a ground state and a first existed state. 
They found that multistate boson stars have higher critical mass than boson stars with the ground state, 
and are more stable than excited cases. 
Moreover, rotating multistate boson stars are constructed and studied~\cite{Li:2019mlk,Li:2020ffy}. 
Besides, other examples are provided here, 
charged boson stars are the scalar field coupled to the electromagnetic field $e.g.$~\cite{Jetzer:1989av,Kleihaus:2009kr,Pugliese:2013gsa}, 
fermion-boson stars consist of bosonic and fermionic matter which is approximately described as perfect fluids $e.g.$~\cite{Henriques:1989ar,Henriques:1989ez,Liebling:2012fv}, 
and newtonian configurations of boson multistate were introduced in~\cite{Blasone:2001du}. 
In addition to this, a lot of interesting studies on BSs have been show in Refs.~\cite{Seidel:1991zh,Kleihaus:2005me,Brito:2015pxa,Herdeiro:2017fhv}, 
and recent studies of boson stars~\cite{Herdeiro:2020kvf,Herdeiro:2021mol,Santos:2020pmh,Kunz:2019sgn,Helfer:2020gui,Jaramillo:2020rsv,Herdeiro:2018wvd,Sanchis-Gual:2019ljs,DiGiovanni:2020ror,Herdeiro:2020kba} have also received a lot of attention. 

Recently, the electron recoil excess observed by XENON1T, which has a possible interpretation in terms of solar axions coupled to electrons~\cite{Aprile:2020tmw}. 
The axion is seen as a plausible candidate for dark matter. 
To solve the strong CP problem~\cite{Peccei:1977hh,Jackiw:1976pf}, the quantum chromo dynamics (QCD) axion is proposed~\cite{Weinberg:1977ma,Wilczek:1977pj}. 
In addition, axion like particles (ALPs) also play an important role in string theory~\cite{Svrcek_2006}. 
D. Guerra, C. F. Macedo and P. Pani~{\em et al.} studied the solutions of spherical symmetry axions boson stars (ABSs)~\cite{Guerra:2019srj}. 
This work was extended to rotating axion boson stars (RABSs)~\cite{Delgado:2020udb,Delgado:2020hwr}. 
Particularly, they suggest that the axion potential reduces to the massive, free, complex scalar field, for $f_a$ is large, 
which matches the solution of the standard spinning mini-boson stars in the context of Einstein gravity coupled to free scalar field~\cite{Schunck:1996he,yoshida1997rotating}. 

The scope of this work is to construct the solutions of rotating hybrid axion-miniboson stars (RHABSs), which is a mixture of bosonic and axionic matter, minimally coupled to gravity respectively. 
By numerical calculation, we study the mass and the angular momentum for different frequency. 
The configuration is multifield and multistate. 
We refer to this model as rotating hybrid axion-miniboson stars (RHABSs). 

This paper is organized as follows: 
In Sec.~\ref{sec2}, we present the model of four-dimensional Einstein gravity minimally coupled to a massive free scalar field and a QCD axion field. 
In Sec.~\ref{sec3}, boundary conditions of the RHABSs are shown. 
In Sec.~\ref{sec4}, we exhibit the numerical results and show the properties of $^1S^2S$ state and $^1S^2P$ state for different axion decay constants $f_a$ and  scalar particle mass $\mu_2$. 
In Sec.~\ref{sec5}, conclusions and perspective are given.

\section{The model setup}\label{sec2}

We consider the case of an axion field and a free complex massive scalar field, which is coupled to $(3+1)$-dimensional Einstein gravity. 
The action is 
\begin{equation}
  S=\int\sqrt{-g}d^4x\left(\frac{R}{16\pi G}+\mathcal{L}_{m}\right) \ ,
\end{equation}
 where first term represents Einstein gravity, 
 Second term ${\cal L}_{m}$ is marked as the matter Lagrangian about an axion field and a scalar filed, 
\begin{equation}
  \label{lag}
  \mathcal{L}_{m}=\sum_{i=1}^2 [-\nabla_a\psi_i^*\nabla^a\psi_i
  - U_i(|\psi_i|^2)] \ .
\end{equation}
Here $\psi_i\ (i=1,2)$ represent the axion field and the scalar field, respectively, 
and $U_i(|\psi_i|^2)$, $(i=1,2)$ are the axion potential and the scalar potential, respectively. 
By varying the action, we can derive the equation of motion. The field equations are
\begin{equation}
  \label{eq:EKG1}
  E_{\mu\nu}=R_{\mu\nu}-\frac{1}{2}g_{\mu\nu}R-8\pi T_{\mu\nu}=0 \ ,
\end{equation}
\begin{equation}
  \label{eq:EKG2}
  \Box\psi_i-\frac{\partial U_i}{\partial |\psi_i|^2}\psi_i=0 \hspace{5pt}, \hspace{5pt} i=1,2 \ .
\end{equation}
Where 
\begin{equation}
  \label{eq:Energy-MomentumTensor}
  T_{\mu\nu}=\sum_{i=1}^2 \left[2\nabla_{(\mu}\psi_i^{\ast}\nabla_{\nu)}\psi_i-g_{\mu\nu}(\nabla^{\alpha}\psi_i^{\ast}\nabla_{\alpha}\psi_i+U_i)\right] \ ,
\end{equation}
is the energy-momentum tensor associated with the axion field and the scalar field. 

We consider the Kerr metric, 
and adopt the ansatz as follows, see {\em e.g.}~\cite{Herdeiro:2015gia,Herdeiro:2014goa}: 
\begin{eqnarray}
  \label{ansatz}
  d s^{2}=-e^{2 F_{0}(r, \theta)} d t^{2}+e^{2 F_{1}(r, \theta)}\left(d r^{2}+r^{2} d \theta^{2}\right)+e^{2 F_{2}(r, \theta)} r^{2} \sin ^{2} \theta(d \varphi-W(r, \theta) d t)^{2} \ ,
\end{eqnarray}
\begin{eqnarray}
  \psi_i=\phi_{i(n)}(r,\theta)e^{i(m_i\varphi-\omega_i t)} \hspace{5pt}, \hspace{5pt} n=0,1,\cdots \hspace{5pt}, \hspace{5pt} m_i=\pm1,\pm2,  \cdots \hspace{5pt}, \hspace{5pt} i=1,2 \ .
  \label{field_ansatz1}
\end{eqnarray}
Where $F_0$, $F_1$, $F_2$, $W$ and $\phi_{i(n)}$ (i=1,2) depend only on the radial distance $r$ and the polar angle $\theta$. 
The sub-indicator $i$ is used to distinguish between the axion field and the scalar field. 
The subscript $n$ represents the principal quantum number. The state with $n=0$ represents the ground state, and the state with $n \geq 1$ is the excited state. 
The $m_i$ and $\omega_i$ are called the azimuthal harmonic index and the frequency of field, respectively. 
The frequency of the field is called the synchronized frequency when $\omega_1=\omega_2=\omega$. 
The case of $\omega_1\neq\omega_2$ is called nonsynchronized frequency. 

Next, we specify the axion field potential and the scalar field potential 
\begin{eqnarray}
  U_1=\frac{2\mu_1^2f_a^2}{B}\left[1-\sqrt{1-4Bsin^2	\left(\frac{\phi_1}{2f_a}\right)}\right] \ ,
  \label{axion_potential}
\end{eqnarray}
\begin{eqnarray}
  U_2=\mu_2^2\psi_2^2 \ .
  \label{scalar_potential}
\end{eqnarray}
Here $B=Z/(1+Z)^2 \approx 0.22$, $Z \equiv \frac{m_u}{m_d} \approx 0.48$, $m_u$ and $m_d$ are mass of the up and down quarks, respectively. 
The axion potential has two free parameters, $\mu_1$ and $f_a$, which is the ALP mass and decay constant, respectively. 
We expand the axion potential around $\phi = 0$, obtaining 
\begin{eqnarray}
U_1(\phi_1)=\mu_1^2\phi_1^2-\left(\frac{3B-1}{12}\right)\frac{\mu_1^2}{f_a^2}\phi_1^4+\cdots \ . \label{expansion}
\end{eqnarray}
Here, $\mu_1$ represents axion mass. $f_a$ is the axion decay constant. 
If $f_a \gg \phi_1$, the axion potential tends to the free scalar potential. 
The model will reduce to the rotating multistate boson stars in Ref.~\cite{Li:2019mlk}.

\section{Boundary conditions}\label{sec3}

In order to solve Eq.~(\ref{eq:EKG1}) and Eq.~(\ref{eq:EKG2}), boundary conditions are necessary. 
The metric functions $F_0(r,\theta)$, $F_1(r,\theta)$, $F_2(r,\theta)$, $W(r,\theta)$ and the field functions $\phi_{i(n)}(r,\theta)$ need to be specified. 
For asymptotically flat solutions, at infinity $r \rightarrow \infty$, the functions must be 
\begin{eqnarray}
F_0=F_1=F_2=W=\phi_{i(n)}=0 \hspace{5pt}, \hspace{5pt} (i=1, 2) \hspace{5pt}, \hspace{5pt}  n=0,1,\cdots \ .
\end{eqnarray}
For axial symmetry, on the axis ($\theta=0,\pi$), 
\begin{equation}\label{abc}
\partial_\theta F_0(r, 0)=\partial_\theta F_1(r, 0)=\partial_\theta F_2(r, 0) =\partial_\theta W(r, 0)=\phi_{i(n)}(r, 0)=0 \hspace{5pt}, \hspace{5pt} n=0,1,\cdots \ .
\end{equation}
For all solutions, we consider the range $\theta \in [0,\pi/2] $ \cite{Herdeiro:2015gia}. 
By specifying the symmetry of matter field, the solutions have two types. 
If matter fields are symmetric for $\theta=\pi/2$ plane, we have, 
\begin{equation}
  \partial_\theta F_0=\partial_\theta F_1=\partial_\theta F_2 = \partial_\theta W=\partial_\theta \phi_{i(n)} = 0 \hspace{5pt}, \hspace{5pt} n=1,2,\cdots \ .
\end{equation}
If matter fields are anti-symmetric for $\theta=\pi/2$ plane, we have, 
\begin{equation}
  \partial_\theta F_0=\partial_\theta F_1=\partial_\theta F_2 = \partial_\theta W= \phi_{i(n)} = 0 \hspace{5pt}, \hspace{5pt} n=1,2,\cdots \ .
\end{equation}
At the origin we require, 
\begin{eqnarray}
  \phi_{i(n)}(0, \theta) = 0 \ ,  \nonumber\\
  \partial_r W(0, \theta) = 0 \ .
\end{eqnarray}
The $F_0(0, \theta), F_1(0, \theta), F_2(0, \theta)$, and $W(0,\theta)$ are unchanged. 

To compute the ADM mass $M$ and angular momentum $J$, we expand $g_{tt}$ and $g_{t\phi}$ at $r\rightarrow\infty$, 
\begin{eqnarray}
\label{asym}
g_{tt}= -1+\frac{2GM}{r}+\cdots \ , \nonumber\\
g_{\varphi t}= -\frac{2GJ}{r}\sin^2\theta+ \cdots \ .
\end{eqnarray}

\section{Numerical results}\label{sec4}
On the one hand, to simplify the form of equations, we use natural units set by $\mu_1$ and $G$, 
\begin{eqnarray}
r \rightarrow r\mu_1 \hspace{5pt}, \hspace{5pt} \phi \rightarrow \phi M_{PI} \hspace{5pt}, \hspace{5pt} \omega \rightarrow \omega/\mu_1 \hspace{5pt}, \hspace{5pt} \mu_2 \rightarrow \mu_2/\mu_1 \ .
\end{eqnarray}
Where $M_{PI}^2=G^{-1}$ is the Plank mass. We set $G = c = \mu_1 = 1$. As a result, $G$ and $\mu_1$ disappear from the equations. 
On the other hand, it's convenient that we transform the radial coordinate $[0,\infty)$ to $[0,1]$, 
\begin{eqnarray}
x=\frac{r}{1+r} \ .
\end{eqnarray}
All numerical calculations based on the finite element methods. The computation has $200 \times 120$ grid points in the integration region $0 \leq x \leq 1$ and 
$0 \leq \theta \leq \pi/2$. 
The relative error for the numerical solutions is less than $10^{-5}$. 

Our work based on rotating boson stars with the first excited state and rotating multistate boson stars. 
Some properties of rotating boson stars deserve to be introduced. 
For the first excited state, we obtain two different types of solutions by setting different boundary conditions~\cite{Wang:2018xhw}. 
When matter field is even-parity for $\theta=\frac{\pi}{2}$, the matter field exists radial nodes which are called $^2S$ state. 
When matter field is odd-parity for $\theta=\frac{\pi}{2}$, the matter field exists angular nodes which are called $^2P$ state. 
For rotating multistate boson stars (RMSBSs), the particles are not all in the same state, but rather existing different states~\cite{Li:2019mlk}. 
In the case of RHABSs, we also don't find this kind of solutions in which bosons are in the same state. 
According to Eq.~(\ref{expansion}), when $f_a \gg \phi_1$, the RHABSs reduce to rotating multistate boson stars (RMSBSs)~\cite{Li:2019mlk,Li:2020ffy}. 
So we follow a similar conception. 
RHABSs composed of a scalar field and an axion field.
We assume that the axion field exists the ground state $^1S$ and the scalar field exists the first excited state $^2S$ or $^2P$. 
If the first excited state is $^2S$ state, scalar field $\phi_2$ with a radial node $n_r=1$, the coexisting state is called as $^1S^2S$ state. 
Besides, if the first excited state is $^2P$ state, scalar field $\phi_2$ with a angular node $n_{\theta}=1$, the coexisting state is called the $^1S^2P$ state. 

According to \cite{Delgado:2020udb}, 
as the axion decay constant $f_{a}$ approaches 0, the results become increasingly complex. 
So the solution of $f_a=0$ cannot be obtained. 
In order to study the properties of the low $f_a$ solutions, 
we choose four different values of the axion decay constant: $f_{a}=\{1.000, 0.025, 0.015, 0.009\}$. 
For simplicity, we only set azimuthal harmonicindex $m_1=m_2=1$. 

\subsection{ \texorpdfstring{$^1S^2S$}{^1S^2S} state}
\begin{figure}[!ht]
	\begin{center}
		\subfigure[]{
			\centering
			\includegraphics[width=3.1 in]{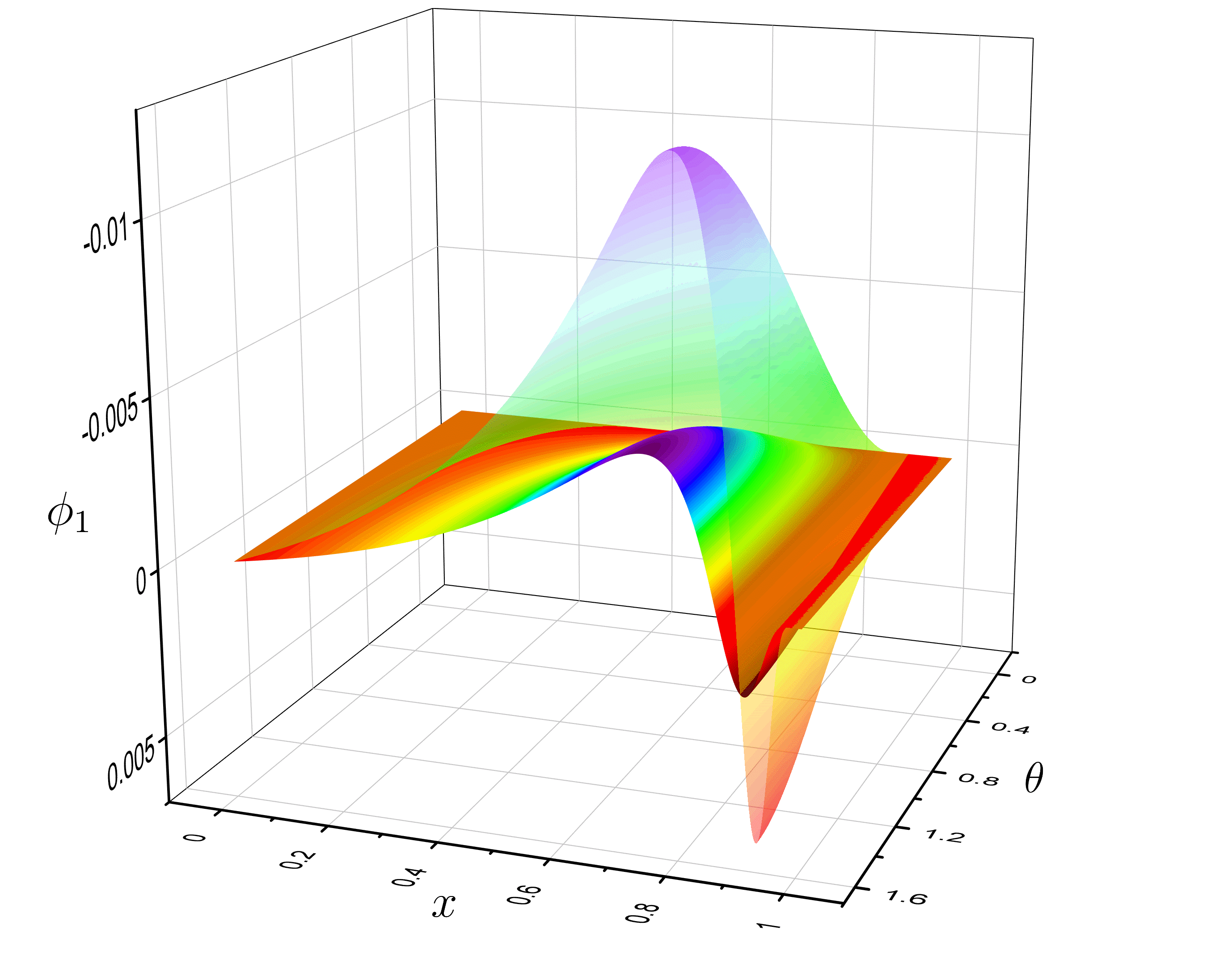}
		}\hspace{-10mm}
		\subfigure[]{
			\centering
			\includegraphics[width=3.1 in]{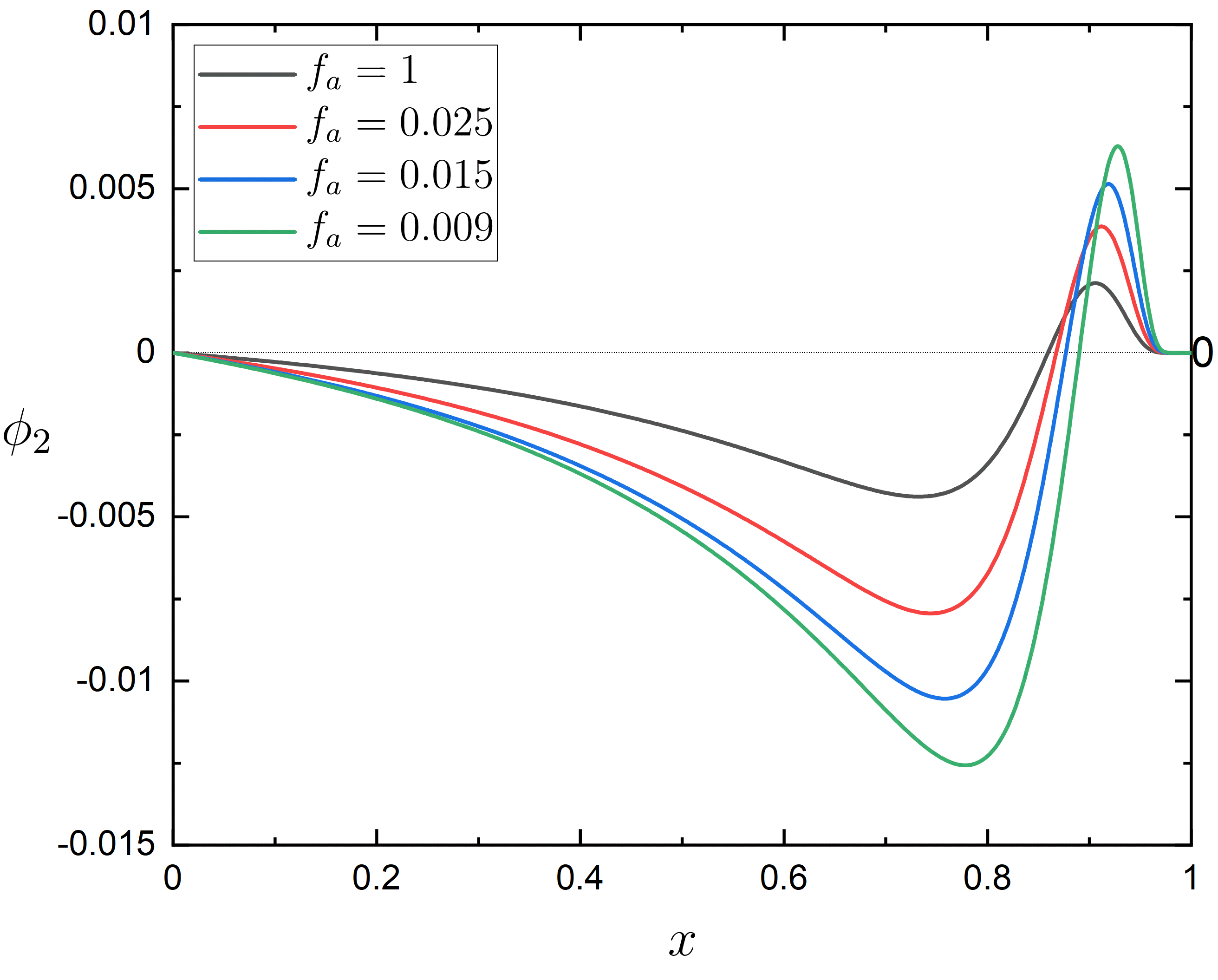}  
		}
	\end{center}
    \caption{\textit{Left}: The distribution of the scalar field $\phi_2$ as a function of $x$ and $\theta$ with the same synchronized frequency $\omega=0.858$. Semi-translucent surface represents $f_a=0.009$ and the opaque surface represents $f_a = 1$. \textit{Right}: At $\theta=\pi/2$, the distribution of the scalar field $\phi_2$ with the same synchronized frequency $\omega=0.858$ for $f_{a}=\{1.000, 0.025, 0.015, 0.009\}$. 
    All solutions have $m_1=m_2=1$, $\mu_1=1$, and $\mu_2=0.93$.} 
    \label{fig:1s2s-phi2}
\end{figure}

The distribution of the scalar field $\phi_2$ for different axion decay constants $f_a$ are presented in Fig.~\ref{fig:1s2s-phi2}. 
In the left panel, we show the two-dimensional distribution of the axion field $\phi_1$. 
Semi-translucent surface indicate $f_a = 0.009$, and the opaque surface indicate $f_a = 1$. 
In the right panel, we plot the distribution of the scalar field $\phi_2$ at $\theta=\pi/2$ for different axion decay constants $f_a$. 
For lower $f_a$, the maxima and minima of the scalar field $\phi_2$ becomes higher. 
Meanwhile, we observer that the scalar field $\phi_2$ changes sign once alone the radial direction $r$. 
In this subsection, the properties of the solutions with even-parity matter field $\phi_1$ and $\phi_2$ will be exhibited. 
Along the angular $\theta$ direction, the values of the axion fields $\phi_1$ and the scalar field $\phi_2$ have no any node. 
Along the radial $r$ direction, the axion field $\phi_1$ remains the same sign, the scalar field $\phi_2$ have a node. 
Meanwhile, to explore the influence of the scalar field mass $\mu_2$ on the RHABSs, 
we show the mass $M$ of the RHABSs versus the synchronized frequency $\omega$ as well as 
the nonsynchronized frequency $\omega_2$ ( $\omega_1=0.8$ ) with the $f_a=\{1.000, 0.025, 0.015, 0.009\}$.
And then, we study the existence domain of the synchronized frequency $\omega$ for different $f_a$ and $\mu_2$. 
The relationship between mass $M$ and angular momentum $J$ will be discussed later in \ref{1s2p}. 


\begin{figure}[ht!]
  \centering
      \includegraphics[width=\textwidth]{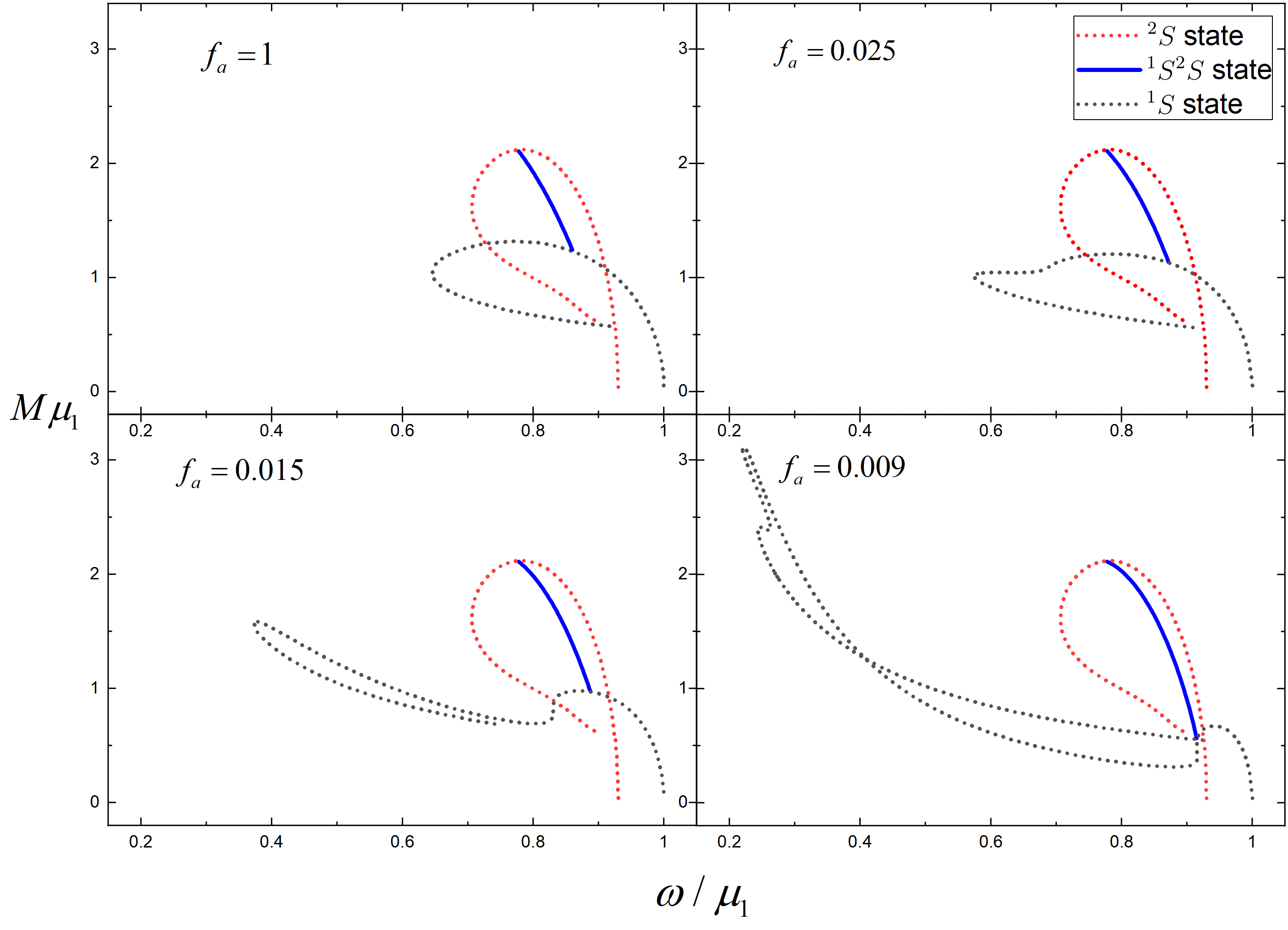}
      \caption{The mass $M$ of  the RHABSs as a function of the synchronized frequency $\omega$ ($\omega_1=\omega_2=\omega$) for $f_a=\{1.000, 0.025, 0.015, 0.009\}$. 
      The black dotted line indicates RABSs with the $^1S$ state, The red dotted line indicates rotating boson stars with the $^2S$ state, and the blue line represents RHABSs, respectively.}
      \label{1s2s-km-synchronized}
  \end{figure}


At the case of synchronized frequency in Fig.~\ref{1s2s-km-synchronized}, 
we study the mass $M$ of the RHABSs versus the synchronized frequency $\omega$ with the $f_a=\{1.000, 0.025, 0.015, 0.009\}$, respectively. 
RHABSs is self-gravitational bound state. 
In order to ensure that RHABSs are bound states, $\omega_i \leq \mu_i$ is necessary. 
Thus, the black dotted line (RABSs with $^1S$ state) starts at $\omega_1=1$, 
and the red dotted line (rotating boson stars with $^2S$ state) starts at $\omega_2=0.93$. 
When $f_a=1$ and $f_a=0.025$, the black dotted line starts at $\omega_1=1$, and as the frequency $\omega_1$ decreases, the mass $M$ gradually increases to a maximum, 
then decreases, reaching an inflection point and entering a second branch, 
in which the mass decreases as the frequency increases; 
When $f_a=0.015$, the black dotted line starts at $\omega=0.93$, and as the frequency decreases, 
the mass gradually increases to a local maximum, 
suddenly decreases to a local minimum, then increases to a global maximum, 
reaching an inflection point and entering an another branch, 
in which the mass decreases as the frequency increases; 
$f_a=0.009$, the curves are more complex and more branches appear. 
The red dotted line starts at $\omega_2=0.93$ and spirals to the center.  
The blue line (RHABSs with coexisting state $^1S^2S$) is sandwiched between the red dotted line and the black dotted line. 
As the synchronized frequency $\omega$ decreases, 
the mass $M$ of the RHABSs increases. 
The coexisting state has a higher mass than the RABSs with $^1S$ state but lower mass than rotating boson stars with $^2S$ state. 
When $f_a=1$, the case is similar to the $^1S^2S$ state of the RMSBSs in Ref.~\cite{Li:2019mlk}. 
The smaller the $f_a$, the greater the difference between RHABSs and RMSBSs. 
We observe that, the axion field vanish when synchronized frequency $\omega$ tends to its minimum, 
and there exists only a single scalar field with the first excited state $^2S$. 
On the contrary, 
the scalar field vanish when synchronized frequency $\omega$ tends to its maximum, 
and there exists only a single axion field with the ground state $^1S$.

\begin{figure}[ht!]
  \centering
      \includegraphics[width=\textwidth]{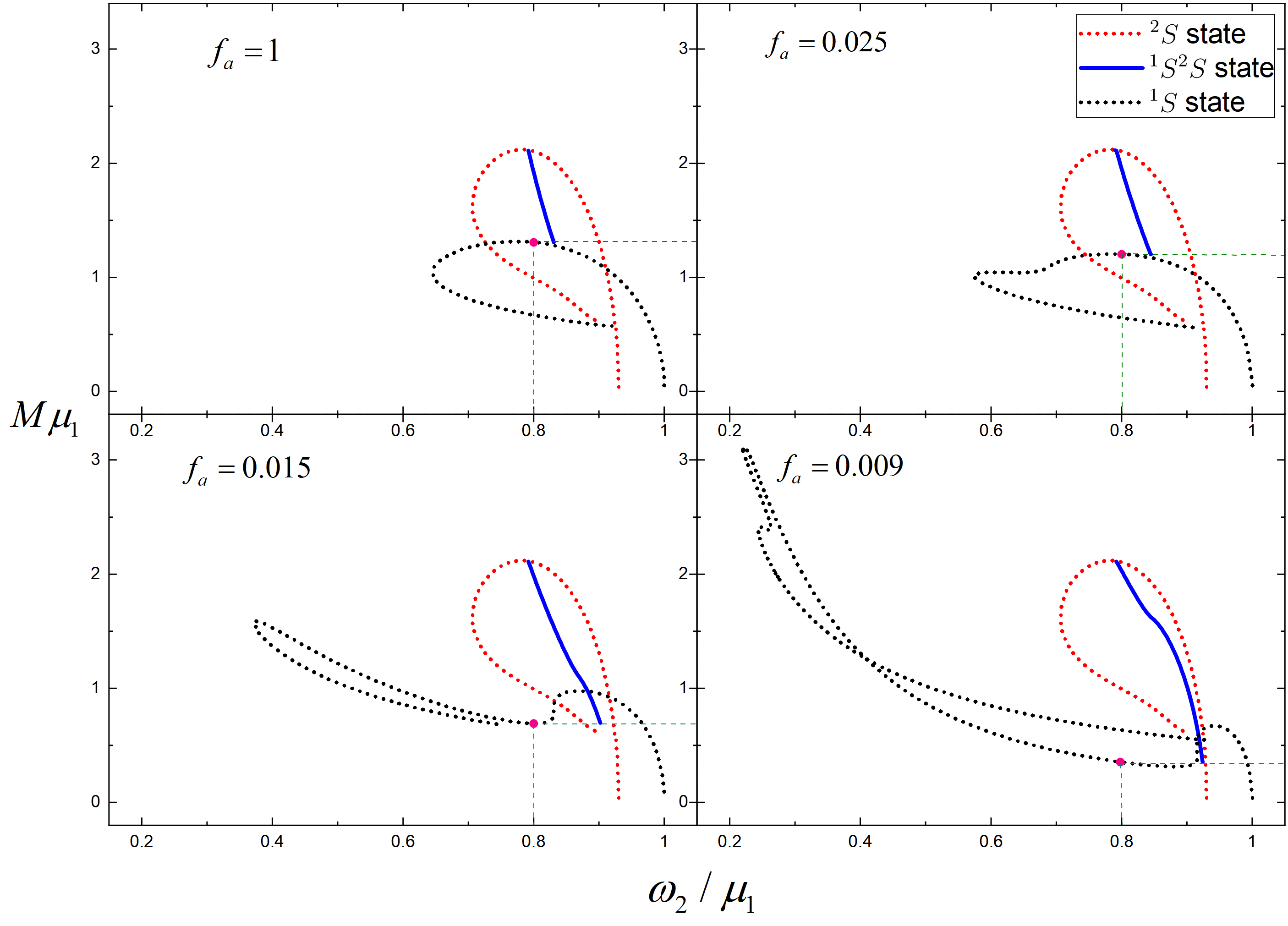}
     \caption{The mass $M$ of  the RHABSs as a function of the nonsynchronized frequency $\omega_2$ ($\omega_1=0.8$) with $f_a=\{1.000, 0.025, 0.015, 0.009\}$ for the fixed parameter $\omega_1=0.8$. 
     The black dotted line indicates the RABSs with the $^1S$ state, The red dotted line indicates rotating boson stars with the $^2S$ state, and the blue line represents RHABSs, respectively. The intersection of the orange lines represents the horizontal and vertical coordinates of the point where the mass $M$ of the RABSs is equal to the minimum mass $M_{min}$ of RHABSs }
    \label{1s2s-km-nonsynchronized}
  \end{figure}

At the case of nonsynchronized frequency, in Fig.~\ref{1s2s-km-nonsynchronized}, 
we show the mass $M$ of the RHABSs versus the nonsynchronized frequency $\omega_2$ with the $f_a=\{1.000, 0.025, 0.015, 0.009\}$. 
Here, we set the frequency of the axion field $\omega_1=0.8$. 
The black dotted line and red dotted line are completely the same as Fig.~\ref{1s2s-km-synchronized}. 
For lower decay constant $f_a$, the blue line RHABSs with nonsynchronized frequency represents is more curved than the blue line RHABSs with synchronized frequency represents in Fig.~\ref{1s2p-km-synchronized}. 
With the nonsynchronized frequency $\omega_2$ increases, the mass $M$ of the RHABSs decreases,  
even the blue blue line crosses the black dotted line for lower $f_a$. 
The fuchsia dot indicates that the mass $M$ of RABSs with $\omega_1=0.8$, which is equal to the minimum mass $M$ of RHABSs. 
Comparing nonsynchronized frequency with synchronized frequency, the minimal mass of RHABSs is lower than the mass of RABSs for $f_a=0.015$ and $f_a=0.009$. 
By analyze fuchsia dot, the mass $M_{min}$ of the RHABSs is provided completely by the axion field, and the scalar field $\phi_2$ tends to $0$. 
Besides, as the decrease of the nonsynchronized frequency $\omega_2$, the $^1S^2S$ state reduces to the first excited state $^2S$. 
From the blue line, we can see that when the synchronized frequency $\omega$ and nonsynchronized frequency $\omega_2$ approaches its maximum, 
the mass is smallest and the RHABSs reduce to the RABSs with the ground state. 

\begin{table*}[t]
  \caption{The existence domain of the synchronized frequency $\omega$ depends on the axion decay constant $f_a$ and the scalar field mass $\mu_2$ for the $^1S^2S$ state. We set $\mu_1=1$ and $m_1=1$.The horizontal axis represents the axion decay constant $f_a$. The vertical axis represents scalar field mass $\mu_2$. \\}
  \centering
  \scalebox{1}{
         \begin{tabular}{|c|c|c|c|c|}
         \hline
  {}&$f_a = 1$&$f_a = 0.025$&$f_a = 0.015$&$f_a = 0.009$\\ \hline
  $\mu_2 = 0.93$&$0.777\sim0.858$&$0.777\sim0.865$&$0.777\sim0.880$&$0.777\sim0.914$\\
  \hline
  $\mu_2 = 0.85$&$0.653\sim0.748$&$0.653\sim0.770$&$0.653\sim0.825$&$0.653\sim0.845$\\
  \hline
  $\mu_2 = 0.80$&$0.608\sim0.698$&$0.608\sim0.720$&$0.608\sim0.780$&$0.608\sim0.794$\\
  \hline
  \end{tabular}}
  \label{1s2s-existence}
  \end{table*}

Table~\ref{1s2s-existence} represents the existence domain of the synchronized frequency $\omega$ depends on the axion decay constant $f_a$ and the scalar field mass $\mu_2$ for the $^1S^2S$ state. 
As the decrease of decay constant $f_a$, the existence domain of the synchronized frequency $\omega$ increases. 
With the scalar field mass $\mu_2$ decrease, the existence domain of the synchronized frequency $\omega$ decreases overall. 
This means that, the smaller $\mu_2$, the more the $M-\omega$ curve shifts to the left.
Since the minimum of $\omega$ represents the RHABSs with only one excited state, 
the axion decay constant has only a very weak effect on the minimum of $\omega$.

\subsection{ \texorpdfstring{$^1S^2P$}{^1S^2P} state}\label{1s2p}

\begin{figure}[!ht]
	\begin{center}
		\subfigure[]{
			\centering
			\includegraphics[width=3.1 in]{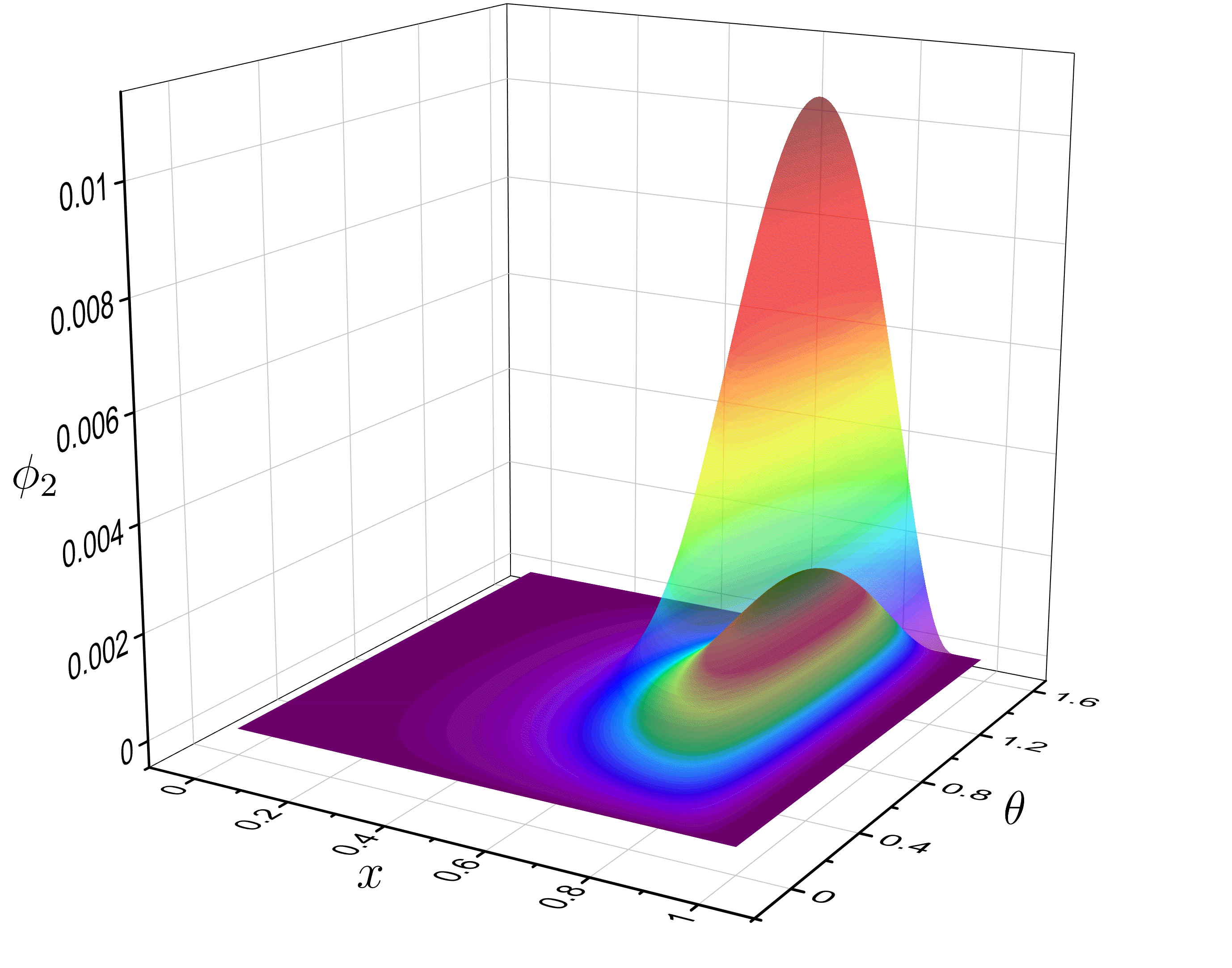}
		}\hspace{-10mm}
		\subfigure[]{
			\centering
			\includegraphics[width=3.1 in]{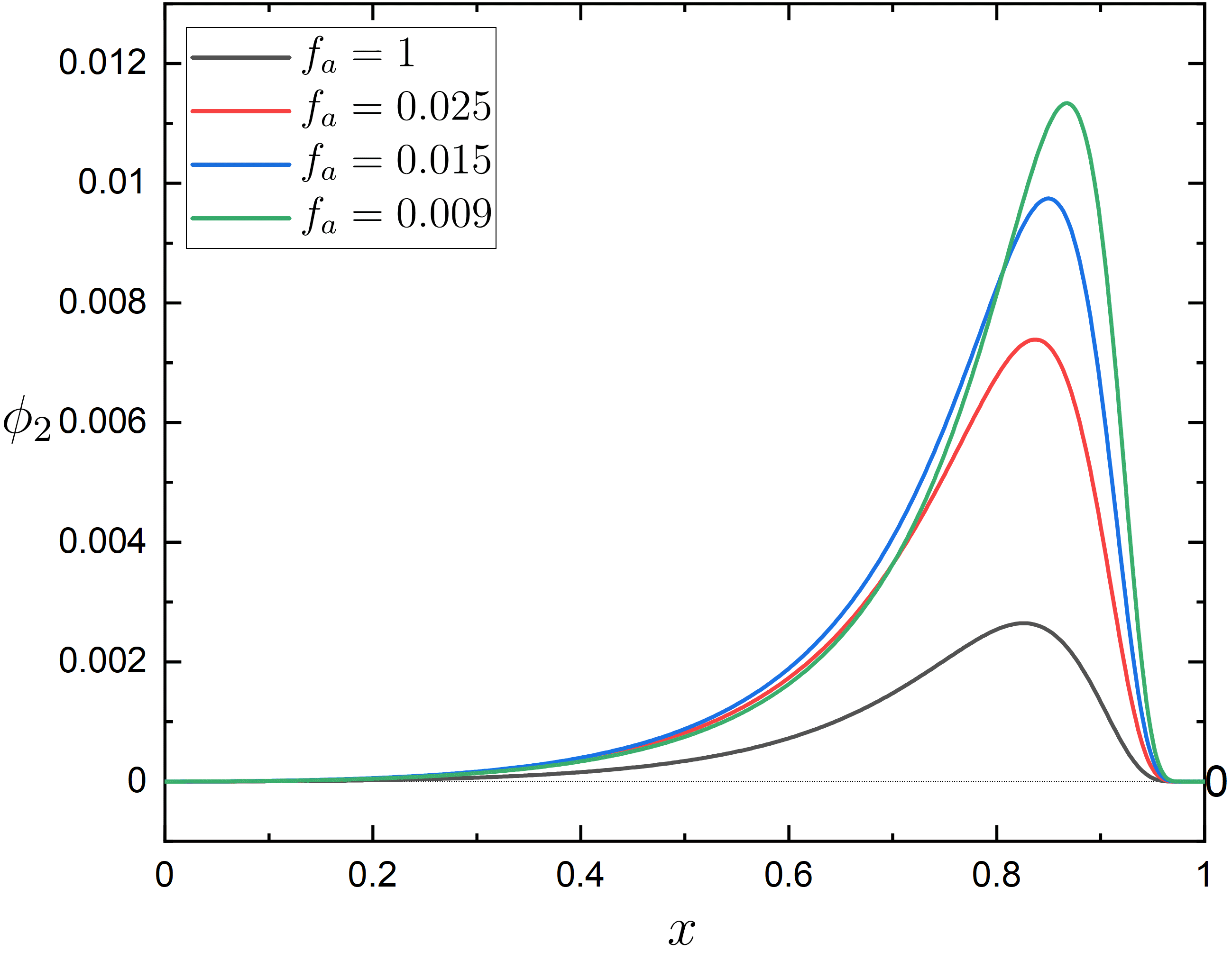}
		}
	\end{center}
  \caption{For $^1S^2P$ state, the distribution of the axion field $\phi_1$ as a function of $x$ and $\theta$ where the axion decay constant $f_a=1$ (left panel) with the same synchronized frequency $\omega=0.842$, and the axion field $\phi_1$ where the axion decay constant $f_a=0.009$ (right panel) with the same synchronized frequency $\omega=0.842$, the scalar field $\phi_2$ as a function of $x$ with $f_a=1$ and $f_a=0.009$ at $\theta=\pi/4$ (bottom panels). All solutions have $m_1=m_2=1$, $\mu_1=1$, and $\mu_2=0.93$.} 
  \label{fig:1s2p-phi2}
\end{figure}

We exhibit the effect of the axion decay constant $f_a$ on the distribution of the scalar field $\phi_2$ in the upper left panel of Fig.~\ref{fig:1s2p-phi2}. 
For lower axion decay constant $f_a$, the maxima of the scalar field $\phi_2$ becomes higher and the distribution of free scalar field becomes more steeper. 

In the last subsection, we gave a family of boson star solutions with even-parity matter field $\phi_1$ and $\phi_2$. 
In this subsection, we will show the properties of the solutions with two odd-parity field $\phi_1$ and $\phi_2$. 
Along the angular $\theta$ and the radial $r$ directions, both the axion field $\phi_1$ and the scalar fields $\phi_2$ keep the same sign (the node exists $\theta=\pi/2$ for the first excited state $\phi_2$). 
This kind of solutions are called $^1S^2P$, which is more unstable than the $^1S^2S$ state~\cite{Wang:2018xhw}. \\

\begin{figure}[ht!]
\centering
    \includegraphics[width=\textwidth]{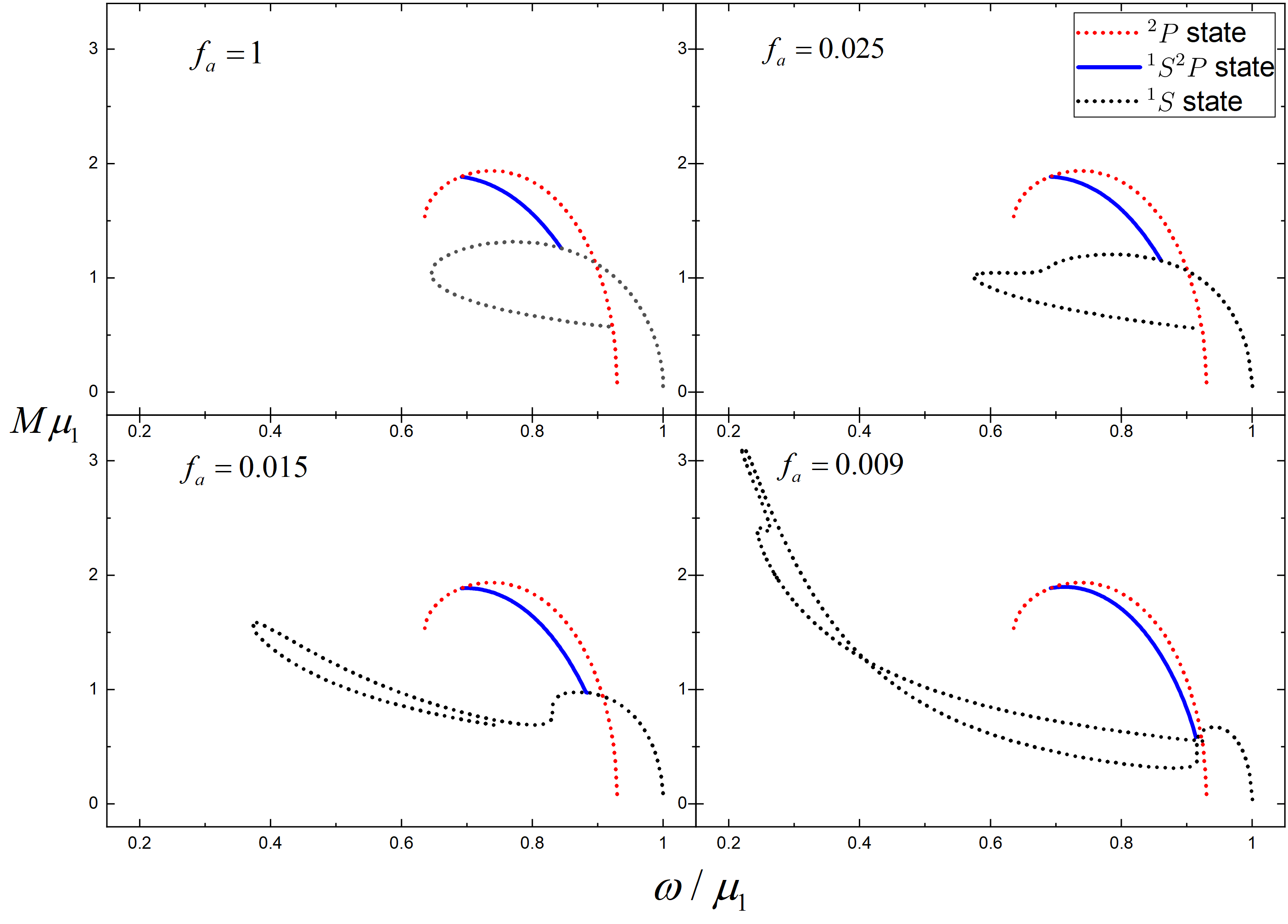}
   \caption{The mass $M$ of  the RHABSs as a function of the synchronized frequency $\omega$ ($\omega_1=\omega_2=\omega$) for $f_a=\{1.000, 0.025, 0.015, 0.009\}$ for the case of $^1S^2P$ state. 
   The black dotted line indicates the RABSs with the $^1S$ state, The red dotted line indicates the rotating boson stars with $^2P$ state, and the blue represents the coexisting state of the ground state and the first excited state, respectively. All the above solutions have $m_2=1$.}
  \label{1s2p-km-synchronized}
\end{figure}

\begin{figure}[ht!]
\centering
    \includegraphics[width=\textwidth]{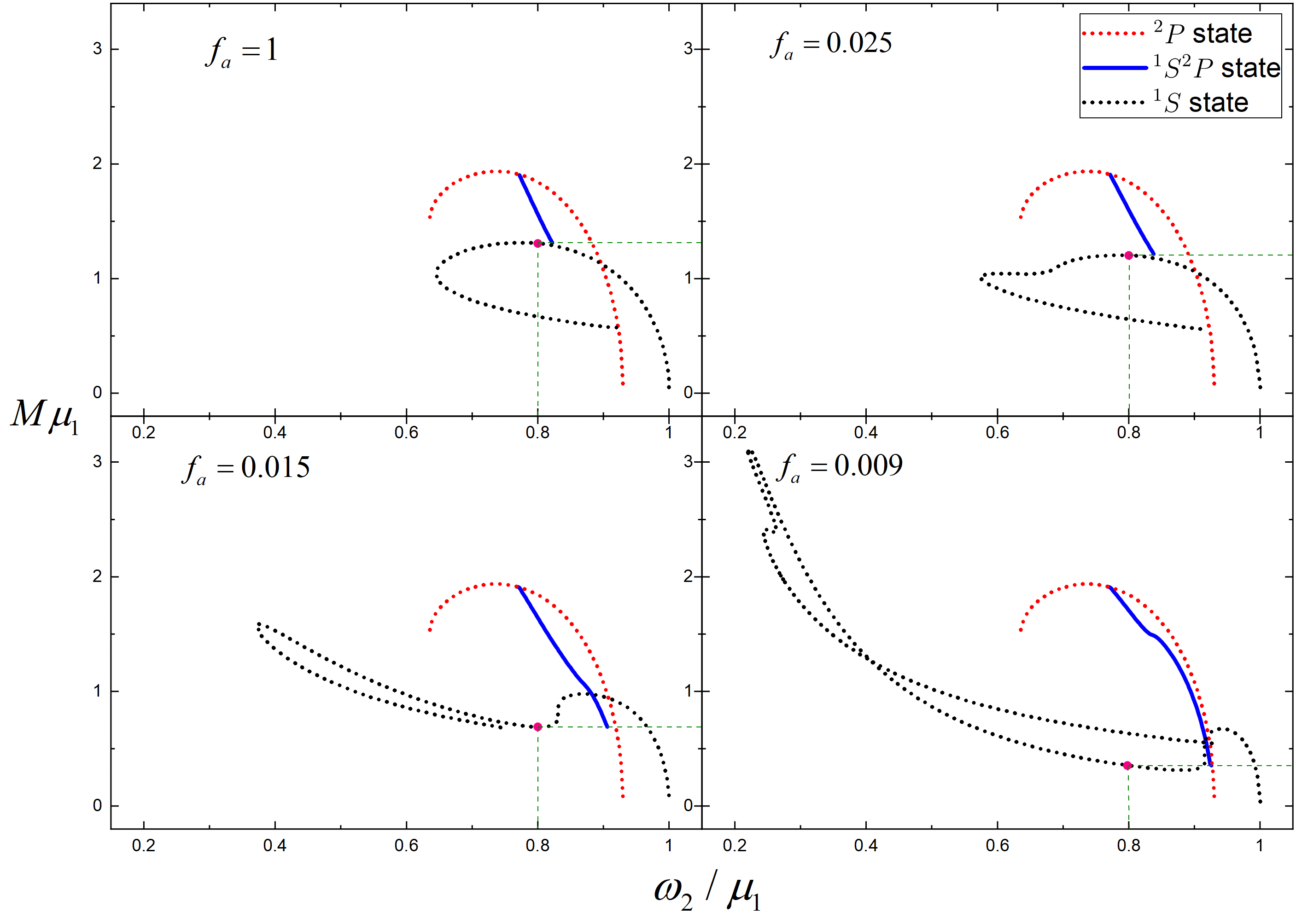}
   \caption{The mass $M$ of  the RHABSs as a function of the nonsynchronized frequency $\omega_2$ ($\omega_1=0.8$) for $f_a=\{1.000, 0.025, 0.015, 0.009\}$ at $^1S^2P$ state. 
   The black dotted line indicates RABSs with $^1S$, The red dotted line indicates rotating boson stars with $^2P$ state, and the blue represents the multistate of the axion field and the scalar field, respectively. All the above solutions have $m_2=1$.}
  \label{1s2p-km-nonsynchronized}
\end{figure}

Based on numerical results, the mass $M$ of the RHABSs as a function of the synchronized frequency $\omega$ as well as the nonsynchronized frequency $\omega_2$ are exhibited in Fig.~\ref{1s2p-km-synchronized} and Fig.~\ref{1s2p-km-nonsynchronized}, respectively. We show the table of the existence domain of the synchronized frequency.  
Furthermore, we discuss the relationship between mass $M$ and angular momentum $J$. 

In Fig.~\ref{1s2p-km-synchronized}, 
The black dotted line (RABSs with the ground state $^1S$) are the same as Fig.~\ref{1s2s-km-synchronized}. 
The red dotted line represents the rotating boson stars with $^2P$ state, which starts at $\omega=0.93$, gradually increases as the frequency decreases, and then decreases as the synchronized frequency decreases. 
The blue line represents RHABSs at the synchronized frequency $\omega$. 
we show the mass $M$ of the RHABSs versus the synchronized frequency $\omega$ with the $f_a=\{1.000, 0.025, 0.015, 0.009\}$. 
the RHABSs exists only a stable branch. 
the mass $M$ of the RHABSs decreases with the synchronized frequency $\omega$ increases. 
When the synchronized frequency $\omega$ increases to maximum, the mass of RHABSs would be minimum. 
The mass of RHABSs is completely provided by RABSs. 
Besides, we note that, as the axion decay constant $f_a$ decreases, 
this maximum value of the synchronized frequency $\omega$ increase, 
and the minimum mass $M$ of the RHABSs decreases. 

In Fig.~\ref{1s2p-km-nonsynchronized}, we show the mass $M$ of the RHABSs versus the nonsynchronized frequency $\omega_2$. 
The black dotted line and red dotted line are the same as Fig.~\ref{1s2p-km-synchronized}. 
The blue line represents RHABSs at the nonsynchronized frequency $\omega_2$, which is more curved than blue line at the case of synchronized frequency. 
the mass $M$ of the RHABSs decreases with the nonsynchronized frequency $\omega_2$ increases. 
When the nonsynchronized frequency $\omega_2$ increases to maximum, the mass of RHABSs would be minimum. It corresponds to the RABSs for $\omega=0.8$. 
The mass of RHABSs is completely equal to the mass of the RABSs for $\omega=0.8$. 
Comparing with the case of synchronized frequency, the blue line crosses the black dotted line. 
Unusually, we find that both $^1S^2S$ state and $^1S^2P$ state possess lower masses than axion field with the ground state at nonsynchronous frequency. 
This is different from the RMSBSs~\cite{Li:2019mlk,Li:2020ffy}. 

\begin{table*}[t]
\caption{The existence domain of the synchronized frequency $\omega$ depends on the axion decay constant $f_a$ and the scalar field mass $\mu_2$ for $^1S^2P$ state. We set $\mu_1=1$ and $m_1=1$.The horizontal axis represents the axion decay constant $f_a$. The vertical axis represents scalar field mass $\mu_2$. \\} 
\centering
\scalebox{1}{
       \begin{tabular}{|c|c|c|c|c|}
       \hline
{}&$f_a = 1$&$f_a = 0.025$&$f_a = 0.015$&$f_a = 0.009$\\ \hline
$\mu_2 = 0.93$&$0.692\sim0.844$&$0.692\sim0.861$&$0.692\sim0.882$&$0.692\sim0.914$\\
\hline
$\mu_2 = 0.85$&$0.584\sim0.750$&$0.584\sim0.776$&$0.584\sim0.828$&$0.584\sim0.846$\\
\hline
$\mu_2 = 0.80$&$0.548\sim0.714$&$0.548\sim0.736$&$0.548\sim0.784$&$0.548\sim0.796$\\
\hline
\end{tabular}}
\label{tab:table2}
\end{table*}

In table~\ref{tab:table2}, the existence domain of the synchronized frequency $\omega$ for different $f_a$ and $\mu_2$ are shown. 
We can see that as the mass $\mu_2$ of the scalar field decreases, 
the existence domain of the synchronized frequency $\omega$ overall decrease. 
This means that the $M-\omega$ curve will shift to the left. 
With the axion decay constant $f_a$ decrease, the upper limit of the synchronized frequency $\omega$ becomes higher for fixed scalar field mass $\mu_2$, the existence domain of the synchronized frequency $\omega$ would be extended. 
Likewise, the minimum of the synchronized frequency $\omega$ almost unchanged. 

\begin{figure}[ht!]
  \centering
    \begin{minipage}[t]{0.47\textwidth}
      \includegraphics[width=\textwidth]{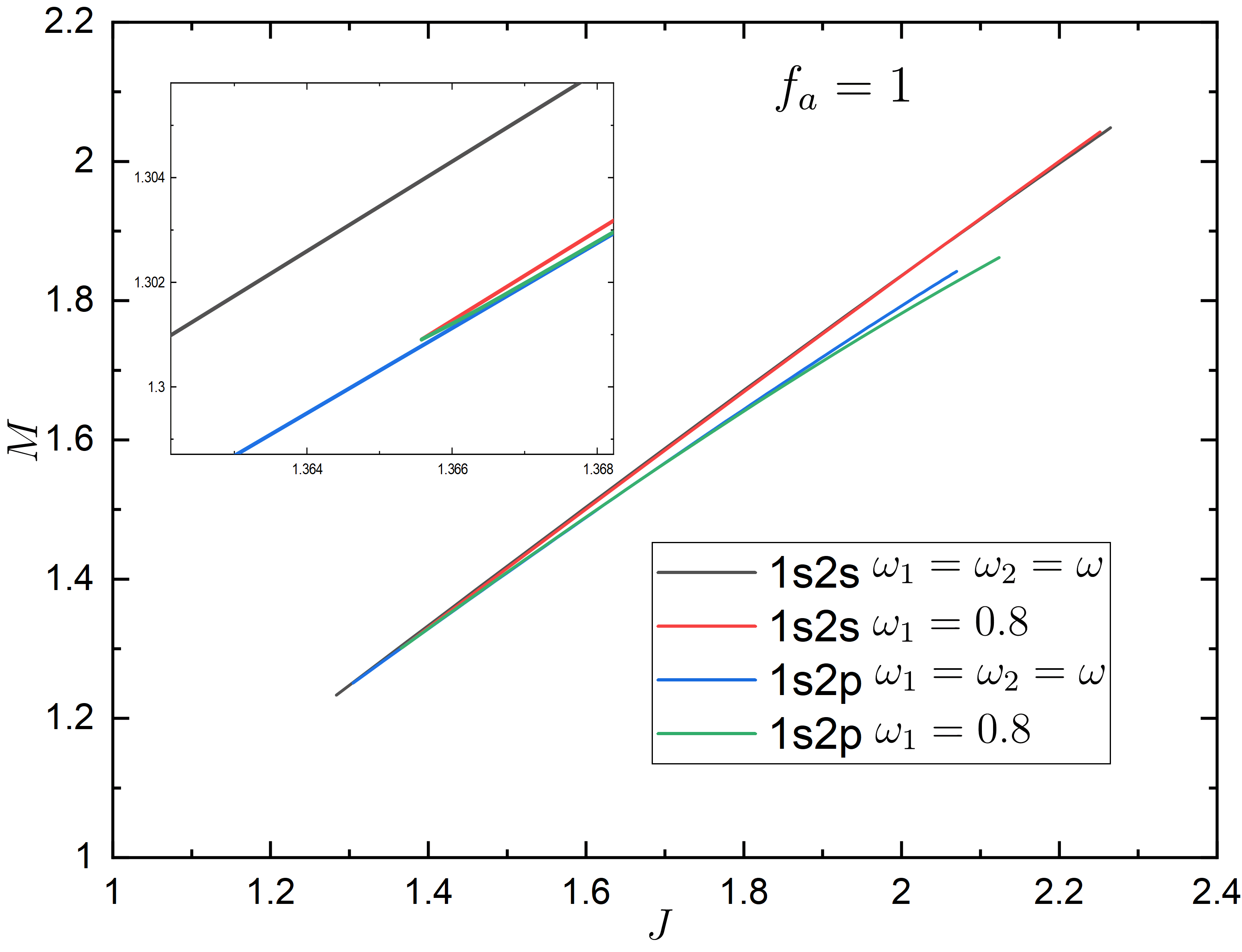}
    \end{minipage}
      \hfill
      \begin{minipage}[t]{0.47\textwidth}
      \includegraphics[width=\textwidth]{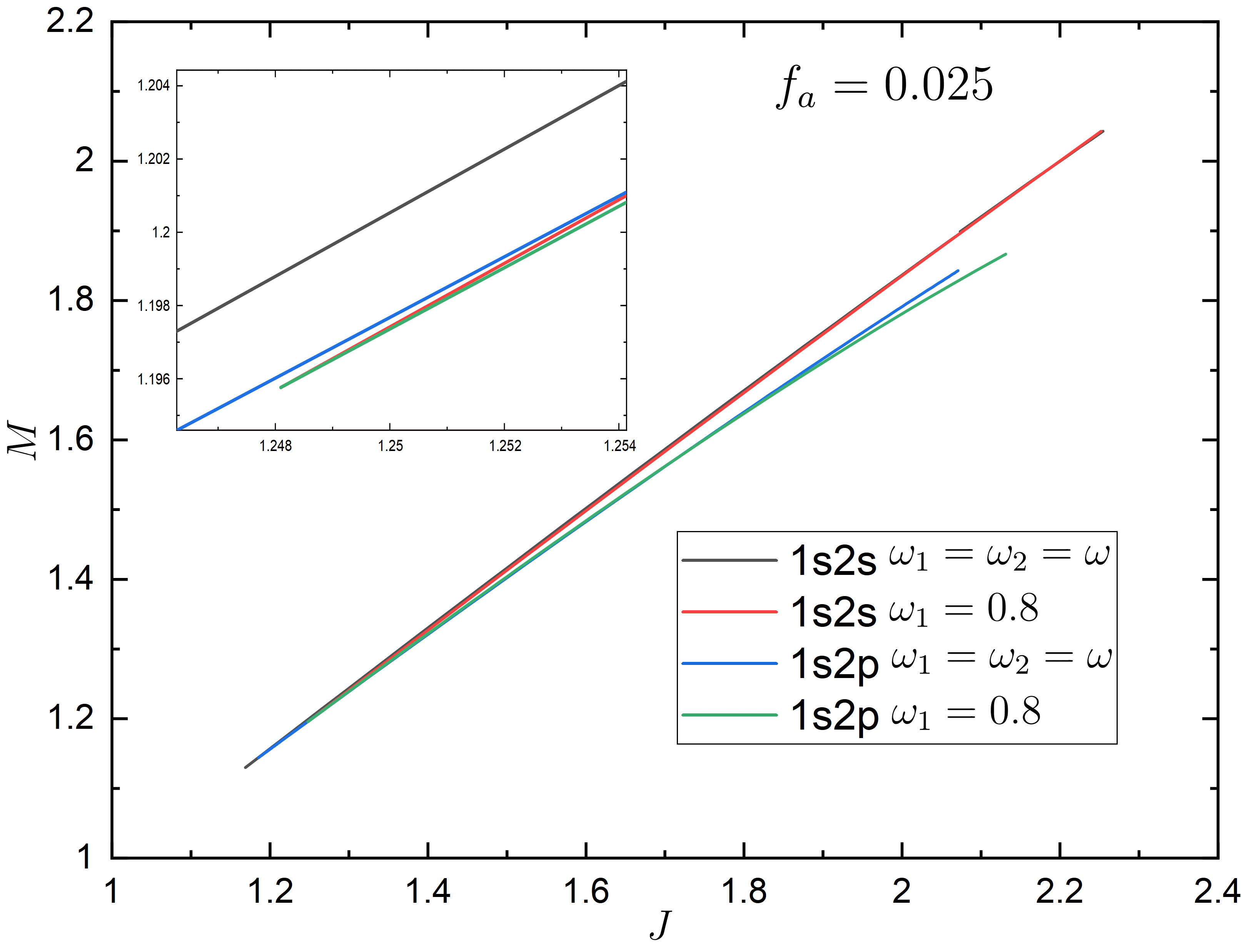}
    \end{minipage}
     \caption{\textit{Left}: The mass $M$ of  the RHABSs versus the angular momentum $J$ for the synchronized frequency $\omega$ and the nonsynchronized frequency $\omega_2$ with $f_a=1$. \textit{Right}: The mass $M$ of  the RHABSs versus the angular momentum $J$ for the synchronized frequency $\omega$ and the nonsynchronized frequency $\omega_2$ with $f_a=0.025$. }
    \label{fig:M-J-1}
  \end{figure}

\begin{figure}[ht!]
    \centering
      \begin{minipage}[t]{0.47\textwidth}
        \includegraphics[width=\textwidth]{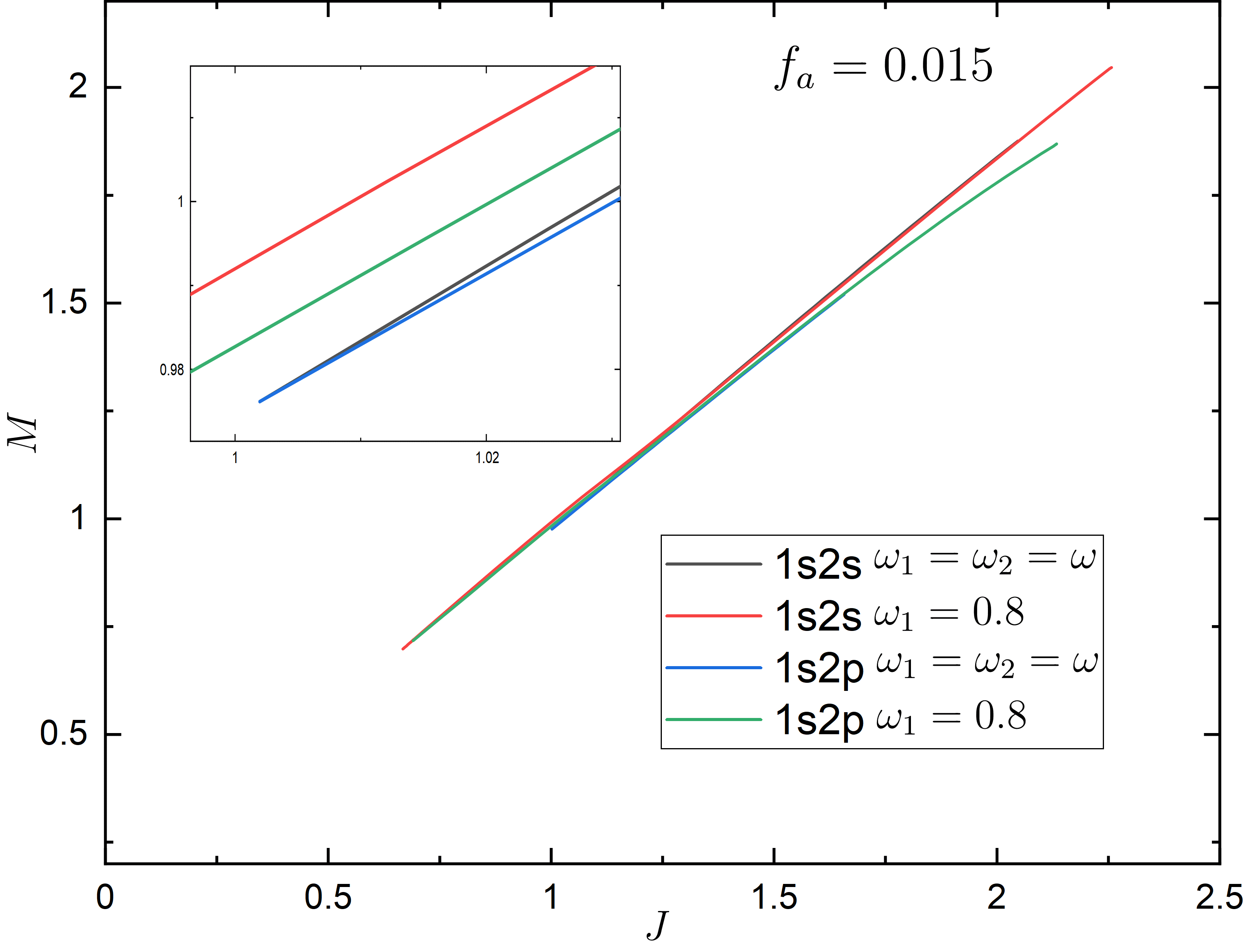}
      \end{minipage}
        \hfill
        \begin{minipage}[t]{0.47\textwidth}
        \includegraphics[width=\textwidth]{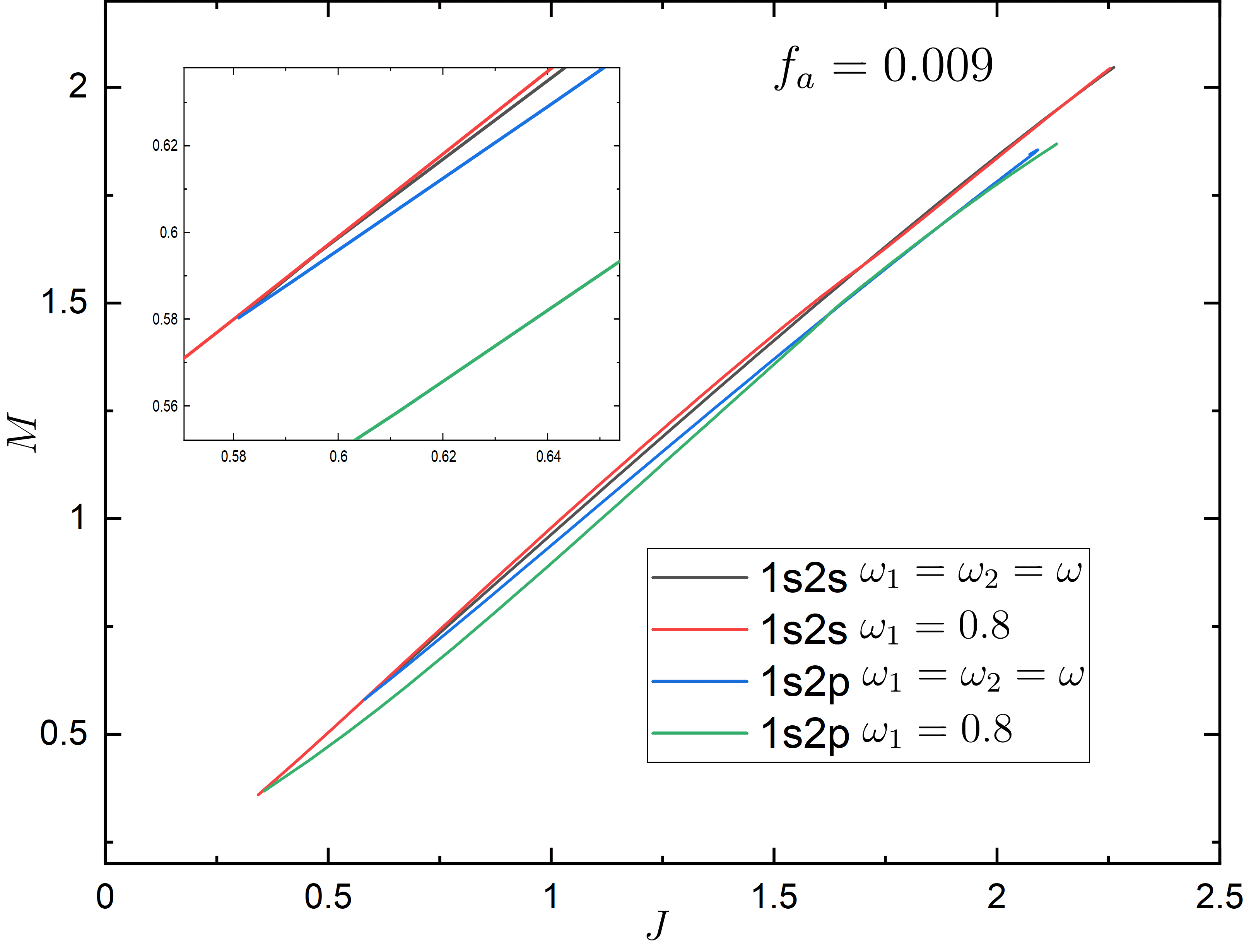}
      \end{minipage}
       \caption{\textit{Left}: The mass $M$ of  the RHABSs versus the angular momentum $J$ for the synchronized frequency $\omega$ and the nonsynchronized frequency $\omega_2$ with $f_a=0.015$. \textit{Right}: The mass $M$ of  the RHABSs versus the angular momentum $J$ for the synchronized frequency $\omega$ and the nonsynchronized frequency $\omega_2$ with $f_a=0.009$. }
      \label{fig:M-J-2}
    \end{figure}

Then, we consider the relationship between the mass $M$ of RHABSs and the angular momentum $J$. 
In Fig.~\ref{fig:M-J-1} and Fig.~\ref{fig:M-J-2}, 
we exhibit the $M-J$ curves for the $^1S^2S$ state and $^1S^2P$ state at synchronized frequency $\omega$ as well as 
nonsynchronized frequency $\omega_2$, when the decay constant of the axion field is respectively equal to $f_a=\{1.000, 0.025, 0.015, 0.009\}$. 
The black dotted line ($^1S^2S$ at synchronized frequency), red dotted line ($^1S^2S$ at nonsynchronized frequency), blue line ($^1S^2P$ at synchronized frequency) and green line ($^1S^2P$ at nonsynchronized frequency) are almost straight line. 
In addition, there is always such feature in the following four graphs. 
The curves of the $^1S^2S$ state and the $^1S^2P$ state at synchronized frequency intersect at one point. 
Similarly, the curves of the $^1S^2S$ state and the $^1S^2P$ state at nonsynchronized frequency intersect at another point. 
The $^1S^2S$ state and the $^1S^2P$ have the same minimum value. 
From the $M - \omega$ curves we can see that when the synchronized frequency and the nonsynchronized frequency approach its maximum, 
the mass $M$ is smallest and RHABSs reduce to the RABSs with the ground state. 
Thus, when the mass $M$ is smallest, 
the $^1S^2S$ state and the $^1S^2P$ state degenerate into RABSs with the ground state $^1S$. 
Comparing the four plots with each other, These diagrams have similar characteristics. 
we can see that as $f_a$ decreases, the minimum value of the angular momentum $J$ and mass $M$ become lower. 

\section{Conclusions}\label{sec5}
In this article, 
we construct a family solutions of rotating boson stars composed of the coexisting states, 
which contains a free scalar field with the first excited state and an axion field with the ground state, 
and analyze the influence of the axion decay constant and scalar particle mass on RHABSs was analyzed. 
Due to the different boundary conditions we chosen, we obtain two types of solutions, 
including $^1S^2S$ state and $^1S^2P$ state. 
We plot the the mass $M$ and the angular momentum $J$ as a function of synchronized frequency $\omega$ and nonsynchronized frequency $\omega_2$, 
and find that the mass $M$ and the angular momentum $J$ decrease with the synchronized frequency $\omega$ and the nonsynchronized frequency $\omega_2$ increase. 
The mass $M$ is positively correlated with angular momentum $J$, and it’s almost proportional.
The RHABSs reduce to a single scalar field with the first excited state as the synchronized frequency and the nonsynchronized frequency approach the minimum, 
the axion decay constant has a very weak effect on the minimum of $\omega$. 
Similarly, the RHABSs reduce to a single axion field with the ground state as the synchronized frequency and the nonsynchronized frequency approach the maximum, 
whether the scalar particles inhabit at $^2S$ or $^2P$ does not affect the RHABSs. 
Thus, We see how the RHABSs degenerates into RABSs with the ground state $^1S$ and rotating boson stars with the first excited state $^2S$ or $^2P$. 
As a result, the $^1S^2S$ state and the $^1S^2P$ state have the same minimal mass $M_{min}$ and the same minimal angular momentum $J_{min}$ regardless of $^1S^2S$ state and $^1S^2P$ state, 
and have the same maximal mass $M_{max}$ and the same maximal angular momentum $J_{max}$ whatever axion decay constant $f_a$ are 
We explore the effect of axion decay constant $f_a$ and scalar field mass $\mu_2$ on the existence domain of the synchronized frequency $\omega$. 
The existence domain of the synchronized frequency $\omega$ overall decrease as the scalar field mass $\mu_2$ decreases, 
and the maximal synchronized frequency $\omega_{max}$ increases as the axion decay constant $f_a$ decreases. 
Moreover, we set four different axion decay constants $f_a=\{1.000, 0.025, 0.015, 0.009\}$. 
When $f_a=1$, $f_a \gg \phi_1$, we can consider RHABSs as RMSBSs.
For enough low decay constant $f_a$, both $^1S^2S$ and $^1S^2P$ possess lower masses than ground state for fixed nonsynchronous frequency $\omega_2$. 
The feature is different from RMSBSs~\cite{Li:2019mlk,Li:2020ffy}. 
In addition, RHABSs has a lower mass limit than RMSBSs as the decay constant $f_a$ decreases. 
    
We will continue to work on the expansion of our research. Firstly, we have studied the RHABSs. 
Next, we will investigate the case of double axion field, 
where one field exists the ground state and the other exists the first excited state. 
This will further study the difference between RMSBSs and RHABSs at the cases of low $f_a$. 
Finally, proca field as massive complex scalar field is very interesting. 
we intend to study the Einstein-complex-proca model and construct the excited Kerr BHs with proca hair in future work.

\section*{Acknowledgements}
YQW would like to thank  Yu-Xiao Liu  and Jie Yang for  helpful discussions. 
Some computations were performed on the  Shared Memory system at Institute of Computational Physics and Complex Systems in Lanzhou University. This work is supported by the National Key Research and Development Program of China Grant No.2020YFC2201503, and the Fundamental Research Funds for the Central Universities (Grants   No. lzujbky-2018-k11 and No. lzujbky-2019-ct06).

\providecommand{\href}[2]{#2}\begingroup\raggedright
\endgroup
\end{document}